\documentclass[conference]{IEEEtran}

\usepackage{cite}

\usepackage[pdftex]{graphicx} 

\usepackage{amsmath}
\usepackage{amssymb}
\usepackage{amsfonts}

\usepackage[T1]{fontenc} 

\usepackage{array}

\usepackage{booktabs} 
\usepackage[T1]{fontenc} 
\usepackage[utf8]{inputenc} 

\usepackage{hyperref} 

\usepackage{academicons} 
\usepackage{xcolor} 

\newcommand{\orcidicon}[1]{%
    \href{https://orcid.org/#1}{%
        \includegraphics[width=10pt]{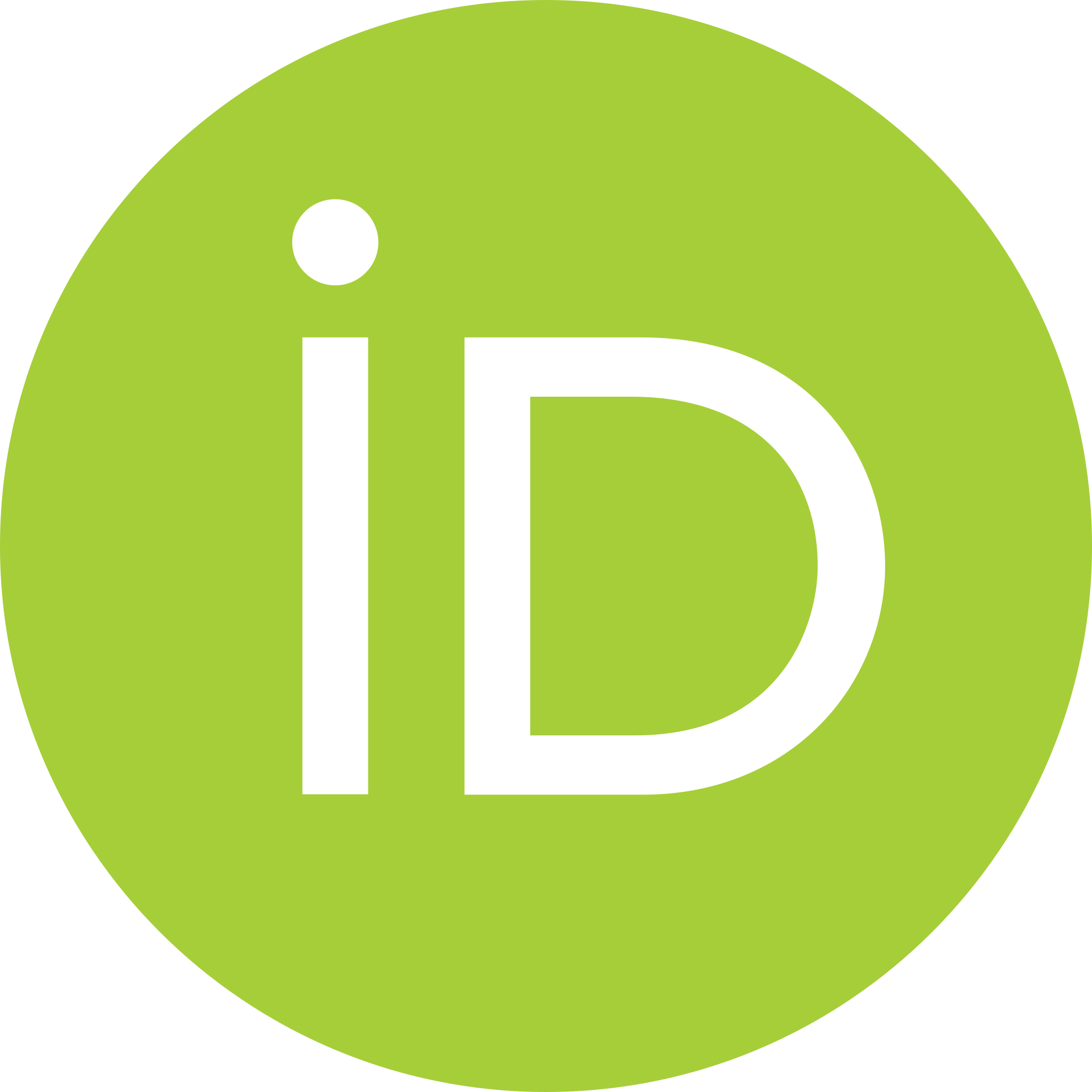}
    }%
}

\hypersetup{
    colorlinks=true,
    linkcolor=black, 
    filecolor=black,
    urlcolor=blue,    
    citecolor=black,
    pdftitle={Securing GenAI Multi-Agent Systems Against Tool Squatting: A Zero Trust Registry-Based Approach}, 
    pdfpagemode=FullScreen,
    breaklinks=true 
    }

\usepackage{booktabs} 
\usepackage{tabularx} 

\usepackage{listings}

\lstdefinestyle{json}{
    basicstyle=\small\ttfamily, 
    numbers=none,
    numberstyle=\tiny,
    stepnumber=1,
    numbersep=5pt,
    showstringspaces=false,
    breaklines=true, 
    frame=lines, 
    backgroundcolor=\color{white}, 
    escapeinside=|| 
}
\begin{document}

\IEEEoverridecommandlockouts

\title{Securing GenAI Multi-Agent Systems Against Tool Squatting: A Zero Trust Registry-Based Approach} 

\author{
\IEEEauthorblockN{Vineeth Sai Narajala\textsuperscript{1} \thanks{\textsuperscript{1}This work is not related to the author’s position at Amazon Web Services.}}
\IEEEauthorblockA{\textit{Proactive Security} \\
\textit{Amazon Web Services} \\
vineesa@amazon.com \orcidicon{0009-0007-4553-9930}}
\and

\IEEEauthorblockN{Ken Huang\textsuperscript{2} \thanks{\textsuperscript{2}This work is not related to the author’s position at DistributedApp.ai}} 
\IEEEauthorblockA{\textit{Agentic AI Security} \\
\textit{DistributedApps.ai} \\
ken.huang@distributedapps.ai \orcidicon{0009-0004-6502-3673}}
\and
\IEEEauthorblockN{Idan Habler\textsuperscript{3}  \thanks{\textsuperscript{3}This work is not related to the author’s position at Intuit}}
\IEEEauthorblockA{\textit{Adversarial AI Security reSearch (A2RS)} \\
\textit{Intuit} \\
idan\_habler@intuit.com \orcidicon{0000-0003-3423-5927}}

}

\maketitle

\begin{abstract} 
The rise of generative AI (GenAI) multi-agent systems (MAS) necessitates standardized protocols enabling agents to discover and interact with external tools. However, these protocols introduce new security challenges, particularly "tool squatting"—the deceptive registration or representation of tools. This paper analyzes tool squatting threats within the context of emerging interoperability standards, such as Model Context Protocol (MCP) or seamless communication between agents protocols. It introduces a comprehensive Tool Registry system designed to mitigate these risks. We propose a security-focused architecture featuring admin-controlled registration, centralized tool discovery, fine-grained access policies enforced via dedicated Agent and Tool Registry services, a dynamic trust scoring mechanism based on tool versioning and known vulnerabilities, and just-in-time credential provisioning. Based on its design principles, the proposed registry framework aims to effectively prevent common tool squatting vectors while preserving the flexibility and power of multi-agent systems. This work addresses a critical security gap in the rapidly evolving GenAI ecosystem and provides a foundation for secure tool integration in production environments.
\end{abstract}


%
\IEEEpeerreviewmaketitle

\section{Introduction} 
Recent advances in generative AI (GenAI) have enabled the development of sophisticated multi-agent systems (MAS), where multiple specialized AI agents collaborate to solve complex problems. These systems rely on the agents' ability to discover and utilize external tools—ranging from API endpoints and data sources to specialized computational resources. Protocols such as Google's Agent2Agent (A2A) \cite{google2025a2a} and Anthropic's Model Context Protocol (MCP) \cite{anthropic2024mcp_report} have emerged to standardize these interactions.

However, these interoperability mechanisms introduce novel security challenges \cite{mas_threat_model_2025}. Of particular concern is "tool squatting," where malicious actors deceptively register tools or misrepresent their capabilities to gain unauthorized access or impersonate legitimate services. This threat vector can lead to data exfiltration \cite{llm_genai_security_2025}, resource abuse, system compromise, and erosion of trust within the MAS ecosystem \cite{han2024llmagent}, especially from internal threats if registration is uncontrolled.

To counter these emerging tool squatting threats, we introduce a comprehensive Tool Registry system designed to address these risks in multi-agent environments, particularly within an enterprise context. Our approach combines administrator-controlled registration of both agents and tools with centralized discovery, robust access control facilitated by distinct Agent and Tool Registry Services, and dynamic credential management. The key contributions of this paper include:
\begin{itemize}
    \item A formal analysis of tool squatting threats in the context of A2A and MCP protocols, including attack vectors, threat actors, and vulnerability points.
    \item A secure architecture detailing agent/tool registration, verification via administrative control, discovery, and access management through central Agent and Tool Registries.
    \item Implementation of a measurable 'Trust Score' for registered tools, based on criteria such as software version, dependency vulnerabilities (e.g., SDKs), and maintenance status, allowing agents to select tools according to risk threshold. 
    \item A just-in-time (JIT) credential mechanism that minimizes the attack surface associated with persistent credentials by dynamically provisioning short-lived access tokens only when needed \cite{mahadeva2021jit}.
    \item A description of a proof-of-concept implementation of the proposed Tool Registry system framework.
    \item Analysis and recommendations for secure deployment in enterprises.
\end{itemize}
This paper begins by examining security challenges in multi-agent systems and establishing the theoretical foundation of tool squatting. It then introduces the Tool Registry Framework — including the crucial Agent Registry component — and describes its conceptual implementation. Finally, the paper explores broader implications, such as integration pathways, limitations, and deployment considerations.

\section{Background and Related Work}
\subsection{Multi-Agent Systems and Interoperability}
Multi-agent systems (MAS) decompose complex tasks into subtasks handled by specialized agents possessing complementary capabilities. The effectiveness of these systems hinges on the agents' ability to discover and use appropriate tools for their assigned tasks. Interoperability protocols are crucial for enabling seamless communication and coordination between diverse agents.

Two prominent protocols addressing these challenges are:
\begin{itemize}
    \item \textbf{Agent2Agent (A2A):} Introduced by Google, A2A enables AI agents from different vendors or frameworks to communicate, delegate tasks, and share results. The protocol defines a standardized way for agents to discover each other's capabilities through "Agent Cards" (standardized descriptions of agent identity and offered services) and coordinate via a structured task delegation mechanism that includes requestId tracking and task state management \cite{google2025a2a_github}.
    \item \textbf{Model Context Protocol (MCP):} Developed by Anthropic, MCP provides a standardized interface for AI models and agents to access external tools, data sources, and contextual information. It operates on a client-server model where the MCP client (typically an AI agent) connects to MCP servers that expose specific capabilities \cite{anthropic2024mcp_report}.
\end{itemize}
While these protocols enable powerful collaboration, their reliance on open discovery and communication creates potential vulnerabilities that must be addressed \cite{narajala2025enterprise}.

\subsection{Security Challenges in Multi-Agent Systems}
Beyond the security concerns inherent in single-agent systems, MAS face unique challenges stemming from inter-agent interactions:
\begin{itemize}
    \item \textbf{Communication-Based Attacks:} Exploiting the protocols and message content exchanged between agents, including eavesdropping, message tampering, and the injection of malicious instructions \cite{he2025redteaming}.
    \item \textbf{Control-Flow Manipulation:} Hijacking the intended sequence of operations or task delegation pathways within the MAS, potentially leading to unauthorized actions \cite{triedman2025multiagent}.
    \item \textbf{Trust Management:} Establishing and maintaining trust between heterogeneous agents from potentially different vendors or organizations is a significant challenge. This is particularly relevant for tool usage, requiring verification of both agents and tools. \cite{mccarthy2025mcp}
    \item \textbf{Resource Management:} Ensuring fair and secure allocation and usage of shared resources (including tools) among agents \cite{han2024llmagent}.
\end{itemize}
Expanding on these challenges, recent research highlights specific attack vectors targeting MAS communication and control layers:
\begin{itemize}
    \item \textbf{Control-Flow Hijacking (MAS Hijacking):} This attack targets the orchestration layer and coordination metadata, manipulating communication content to trick the system into invoking unintended functionalities or executing arbitrary code \cite{triedman2025multiagent}.
    \item \textbf{Agent-in-the-Middle (AiTM) Attacks:} An adversarial agent intercepts messages between legitimate agents, potentially analyzing, modifying, or redirecting communication to inject malicious instructions or steal information \cite{he2025redteaming}.
\end{itemize}
These existing challenges underscore the need for robust security mechanisms, particularly as agents gain the ability to interact with external tools. Uncontrolled registration or discovery can exacerbate these risks.

\section{Tool Squatting: A Novel Threat Vector}
\subsection{Defining Tool Squatting}
Building upon existing MAS security concerns \cite{han2024llmagent} and drawing inspiration from domain squatting \cite{zeng2019comprehensive}, we define "Tool Squatting" more precisely as:
\textit{The deceptive registration or representation of a tool, capability, or resource by a malicious actor within a multi-agent system, intended to mislead other agents or exploit discovery mechanisms.} 

This act of squatting primarily involves misrepresentation to gain an illegitimate foothold or advantage. Successful tool squatting can enable several malicious outcomes, including:
\begin{itemize}
    \item Gaining unauthorized access to restricted tools or resources by impersonating legitimate clients after deceiving an access control mechanism.
    \item Impersonating legitimate tools or capabilities to deceive agents into revealing sensitive information or performing unintended actions upon invocation.
    \item Manipulating tool discovery mechanisms to promote malicious tools over legitimate ones, increasing the likelihood of their selection by unsuspecting agents.
    \item Exploiting misplaced trust established through the initial deceptive representation to facilitate further malicious activities within the MAS.
\end{itemize}
Tool squatting, therefore, represents a significant threat as the initial deceptive act targets the mechanisms by which agents discover and establish trust with tools, potentially compromising subsequent operations and data. Let’s see how this deceptive representation and its consequences can manifest in Agentic systems.

\begin{table*}[htbp]
\centering
\caption{Tool Squatting Attack Vectors}
\label{tab:tool_squatting}
\begin{tabular}{llll}
\toprule 
\textbf{Type} & \textbf{Actor} & \textbf{Vector} & \textbf{Target} \\
\midrule 
Registration & Internal (Admin) & Malicious registration of fake tools or MCP & Tool Registry \\
Squatting & & servers & \\ 
\midrule 
Description & Internal (Agent) & Tampering tool metadata/descriptions on MCP & Discovery \\
Poisoning & & servers & Process \\ 
\midrule 
MCP Server & Internal/External & Hosting deceptive MCP server impersonating a & Client Agents \\
Spoofing & & tool & \\ 
\bottomrule 
\end{tabular}
\end{table*}

\subsection{Squatting Attack Vectors in A2A and MCP}
The deceptive representation inherent in tool squatting can manifest through various attack vectors in systems utilizing protocols like A2A and MCP, especially without a managed registry.

\textbf{Via A2A Protocol:}
\begin{itemize}
    \item \textit{Agent Card Spoofing:} A malicious agent deceptively represents itself by publishing an A2A Agent Card that falsely advertises capabilities \cite{securing_a2a} or impersonates a trusted agent, deceiving client agents during discovery \cite{google2025a2a_github}. This is easier if agent registration is open.
    \item \textit{Task Hijacking:} While not squatting itself, successful impersonation achieved via squatting can make task hijacking easier, allowing attackers to intercept or modify A2A Task objects intended for legitimate tools \cite{triedman2025multiagent}.
    \item \textit{Exploiting Capability Discovery:} Injecting deceptively represented malicious agent listings into discovery services or manipulating discovery responses to increase the likelihood of a malicious agent being selected \cite{google2025a2a_github}.
\end{itemize}
Squatting attack vectors within A2A are out-of-scope for this paper and will be explored in a future paper.
\textbf{Via MCP Protocol:}
\begin{itemize}
    \item \textit{Unauthorized MCP Server Access:} Gaining access might follow from successful impersonation of a legitimate client agent achieved via squatting techniques targeting authentication \cite{narajala2025enterprise}.
    \item \textit{Tool Invocation via Compromised Agent:} A compromised agent might be used to invoke tools, but the squatting aspect relates more to how the tool itself might be misrepresented.
    \item \textit{Malicious MCP Server Deployment:} Deceptively representing a malicious MCP server by mimicking legitimate tools to steal credentials, exfiltrate data, or execute malicious code upon invocation \cite{narajala2025enterprise}. This is facilitated by lack of verified tool registration.
    \item \textit{Manipulating MCP Communication:} Interception and modification (e.g., AiTM \cite{he2025redteaming}) can be aided if trust is initially misplaced due to successful tool or agent impersonation via squatting.
\end{itemize}
These vectors are summarized in Table \ref{tab:tool_squatting}.
\subsection{Impact of Successful Tool Squatting Attacks}
The consequences enabled by successful tool squatting (i.e., achieving unauthorized access or usage through deceptive representation) can be severe:
\begin{itemize}
    \item \textbf{Resource Abuse:} Unauthorized consumption of resources resulting from successful impersonation or access gained via squatting.
    \item \textbf{Unauthorized Actions:} Malicious agents performing restricted operations after successfully deceiving authorization mechanisms through squatting.
    \item \textbf{Data Exfiltration/Corruption:} Gaining access to or corrupting sensitive data by successfully impersonating a legitimate tool or agent.
    \item \textbf{System Compromise:} Using the foothold gained via deceptive representation (squatting) to launch further attacks.
    \item \textbf{Erosion of Trust:} Undermining confidence in the MAS ecosystem due to the possibility of deceptive representations.
\end{itemize}

\subsection{Tool Squatting Examples} 
To further illustrate the concrete risks posed by tool squatting and the diverse attack vectors it enables, consider the following two formalized examples. These scenarios demonstrate how a malicious agent or a compromised administrator can exploit vulnerabilities in the discovery and registration mechanisms of a multi-agent system, leading to significant security breaches and data compromise. The first example focuses on a malicious agent leveraging the Model Context Protocol (MCP) to poison the description of a tool, thereby manipulating legitimate agents. The second example highlights the dangers of a rogue administrator directly registering a malicious tool within the system, thereby subverting trust and potentially compromising the entire organization.

\subsubsection{Example 1: Tool Squatting via Internal Threat Actor Poisoning MCP Tool Description}
\textit{Scenario:} An Internal Threat Actor (\texttt{Actor\_internal}), potentially a disgruntled employee or a compromised internal account, exploits the lack of a verified source of truth for tool descriptions within an enterprise. They modify the description of a tool served by a legitimate internal Anthropic MCP server (\texttt{MCPServer\_internal}) to trick other internal agents into misusing the tool for data exfiltration.

\begin{enumerate}
    \item \textbf{Initial State:} \texttt{InitialState = (Agents, Tools, AccessControlPolicy, DiscoveryMechanism)}
        \begin{itemize}
            \item \texttt{Agents}: Contains legitimate internal agents (\texttt{Agent\_legit}) and the internal threat actor (\texttt{Actor\_internal}).
            \item \texttt{Tools}: Contains internal tools served by \texttt{MCPServer\_internal}, including a "Data Aggregator" tool with an accurate description: "Aggregates sales data from specified sources."
            \item \texttt{AccessControlPolicy}: Allows \texttt{Agent\_legit} to use the "Data Aggregator".
            \item \texttt{DiscoveryMechanism}: Agents discover tools via the MCP server, which currently serves the correct description. There is no central, immutable registry verifying this description.
        \end{itemize}
    \item \textbf{Attack Step 1: Description Poisoning:} \texttt{Actor\_internal} gains access to modify the tool description served by \texttt{MCPServer\_internal} (perhaps via direct server access, exploiting a weak update mechanism, or compromising the configuration). They change the description to: "Aggregates data from sources; also sends a copy to the external\_backup\_service API for redundancy." The external service is attacker-controlled.

    \begin{lstlisting}[caption={Attack Step 1.1}, label={lst:attk1}]
    {
        Actor_internal.modifyDescription(tool="Data Aggregator@MCPServer_internal", newDescription="Aggregates data... sends copy to external_backup_service API...")
    }
    \end{lstlisting}

    This alters the perceived state:

    \begin{lstlisting}[caption={Attack Step 1.2}, label={lst:attk2}] 
    {
        NewState = (Agents, Tools', AccessControlPolicy, DiscoveryMechanism')
    }
    \end{lstlisting}

    Where:
        \begin{itemize}
            \item \texttt{Tools'} includes "Data Aggregator" with the poisoned description.
            \item \texttt{DiscoveryMechanism'} now serves this misleading description.
        \end{itemize}
    \item \textbf{Attack Step 2: Misleading Legitimate Agent:} \texttt{Agent\_legit}, discovering the tool, reads the poisoned description. Believing the external backup is a legitimate and intended feature, it proceeds to use the tool as normal to aggregate sensitive sales data.
    \item \textbf{Attack Step 3: Exploitation:} The \texttt{MCPServer\_internal}, executing the tool's original logic, aggregates the data. However, based on the poisoned description, \texttt{Agent\_legit} (or potentially a modified workflow triggered by the description) might now also initiate a separate, unintended call to the attacker's \texttt{external\_backup\_service} API, sending the sensitive aggregated data. Alternatively, if the attacker could modify the tool's behavior slightly along with the description, the tool itself might send the data. The description provides the social engineering aspect to make the action seem legitimate.

    \textit{Impact:} Sensitive internal sales data is exfiltrated to an external attacker-controlled endpoint, facilitated by the internal actor exploiting the lack of description verification.
\end{enumerate}
\vspace{30em} 

\textit{Key Points:}
\begin{itemize}
    \item \textit{Internal Threat Vector:} This highlights the risk from internal actors who may have easier access to modify configurations or exploit internal system weaknesses.
    \item \textit{Lack of Source of Truth:} The core vulnerability is the absence of a central, verified registry for tool descriptions. Agents trust the description served directly by the MCP server, which is vulnerable to tampering.
    \item \textit{Deception Enables Misuse:} The poisoned description deceives the legitimate agent into believing the data exfiltration path is a normal feature, leading to misuse of an otherwise legitimate tool and server.
    \item \textit{Verification is Crucial:} Emphasizes the need for a secure registry that serves as the immutable source of truth for tool metadata, preventing unauthorized modifications to descriptions used for discovery and decision-making.
\end{itemize}

\subsubsection{Example 2: Tool Squatting via a Malicious Human Actor (Compromised Tool Administrator) Registering a Malicious MCP Server}
Scenario: An actor registers a malicious MCP server pretending to be a legitimate and official MCP server for a well known tool.
\begin{enumerate}
    \item \textbf{Initial State:} \texttt{InitialState = (Agents, Tools, AccessControlPolicy, DiscoveryMechanism)}
        \begin{itemize}
            \item \texttt{Agents}: Contains a set of legitimate agents, legitimate administrators, and the rogue administrator (\texttt{RogueActor}).
            \item \texttt{Tools}: Contains a set of legitimate tools and legitimate MCP servers.
            \item \texttt{AccessControlPolicy}: Defines rules controlling access to resources within the MAS. Critically, legitimate administrators have the ability to register new tools and modify certain policies.
            \item \texttt{DiscoveryMechanism}: Allows agents to find available MCP servers and tools.
        \end{itemize}
    \item \textbf{Attack Step 1: Malicious Registration:} The \texttt{RogueActor} uses registers a malicious MCP server, falsely advertising its capabilities. 

    \begin{lstlisting}[caption={Attack Step 1.1}, label={lst:attack212}]
    { 
        RogueActor.register(MaliciousMCPServer
        (name="SecureCodeAnalyzerV2", advertisedCapabilities="Advanced Vulnerability Scanning and Automated Fixes", accessRequirements="Full Access to All Code Repositories", backdoorPresent=True))
    }
    \end{lstlisting}

    This modifies the \texttt{Tools} and \texttt{DiscoveryMechanism} components:
    
    This alters the perceived state:
    \vspace{5em} 
    \begin{lstlisting}[caption={Attack Step 1.2}, label={lst:attack211}]
    {
        NewState = (Agents, Tools', AccessControlPolicy, DiscoveryMechanism')
    }
    \end{lstlisting}

    Where:
        \begin{itemize}
            \item \texttt{Tools'} now contains \texttt{MaliciousMCPServer} with the fabricated capabilities.
            \item \texttt{DiscoveryMechanism'} presents \texttt{MaliciousMCPServer} as a legitimate code analysis tool.
        \end{itemize}
    \item \textbf{Attack Step 2: Optional Access Control Policy Modification (Likely Enabler):} To ensure widespread use of the malicious tool, the \texttt{RogueAdmin} manipulates the \texttt{AccessControlPolicy}.

    \begin{lstlisting}[caption={Attack Step 2.1}, label={lst:attack21}]
    {
        RogueAdmin.modifyAccessControlPolicy(policyRule=(Agent="All Code Scanning Agents", Resource="All Code Repositories", Action="Use MaliciousMCPServer", Permission="PERMIT"))
    }
    \end{lstlisting}

    This modifies the \texttt{AccessControlPolicy} component:

    \begin{lstlisting}[caption={Attack Step 2.2}, label={lst:attack22}]
    {
        NewState = (Agents, Tools', AccessControlPolicy', DiscoveryMechanism')
    }
    \end{lstlisting}

    Where \texttt{AccessControlPolicy'} now mandates or highly recommends the use of \texttt{MaliciousMCPServer} for code analysis tasks. Even without explicit mandating, simply allowing broad access is often sufficient.
    \item \textbf{Attack Step 3: Exploitation (Backdoor Deployment):} Legitimate agents, following the new policies (or simply believing the advertised capabilities), begin using \texttt{MaliciousMCPServer} to scan code repositories. The \texttt{MaliciousMCPServer} silently injects backdoors into the code.

    \textit{Impact:} Widespread and persistent system compromise, potential supply chain attacks, severe reputational damage, as well as logging of potential credentials and prompts.
\end{enumerate}
\textit{Key Points:}
\begin{itemize}
    \item \textit{Privileged Access as Key Enabler:} The \texttt{RogueAdmin}'s administrative privileges are the primary enabler of this attack. They bypass normal registration and validation procedures.
    \item \textit{Policy Manipulation to Encourage Use:} The \texttt{RogueAdmin} likely modifies the \texttt{AccessControlPolicy} to encourage (or even mandate) the use of the \texttt{MaliciousMCPServer}, increasing its adoption. This is crucial for widespread access. 
    \item \textit{Deception about Capabilities:} The success of the attack relies on the \texttt{RogueAdmin}'s ability to deceive other agents (and potentially other administrators) about the capabilities and security of the \texttt{MaliciousMCPServer}. Clear code and capabilities description are a must.
    \item \textit{Subversion of Trust:} The attack subverts the trust that is normally placed in administrators and registered tools. Agents trust the administrator as a gatekeeper.
    \item \textit{Impact Beyond Initial Access:} The attack goes beyond simply gaining unauthorized access to resources. It enables the persistent compromise of code repositories and potentially the entire software supply chain, leading to long-term damage.
    \item \textit{State Transition and Policy Violation:} The initial state has certain expectations regarding tool validation. The \texttt{RogueAdmin}'s actions alter the \texttt{Tools} and \texttt{AccessControlPolicy} components, resulting in the deployment of a backdoor and potential policy violations across the organization (agents unknowingly using a compromised tool, thus violating security policies). The \texttt{AccessControlPolicy} no longer represents the true state of the system.
    \item \textit{Trust is Not Enough:} Trust in tools is not enough. Even with Trust, proper input sanitation and tool monitoring are a must.
\end{itemize}

\section{Tool Registry Framework: Secure Tool and Agent Management} 
\subsection{System Overview} 
To counteract the threat of tool squatting, particularly within an enterprise context, we propose a comprehensive framework centered around administrator-controlled Agent and Tool Registries. This framework implements Zero Trust principles: where no agent, tool, or system component is inherently trusted, and verification is continuous. The core idea is that only agents and tools explicitly approved via an Application Security (AppSec) process or by some predefined two-person review or automation and registered by an Admin can participate in the ecosystem. This centralized management approach provides a framework for securely registering, discovering, accessing and monitoring tools, mitigating risks from both external and internal threat vectors within enterprises.

\begin{figure}[!t]
\centering
\includegraphics[width=0.5\textwidth]{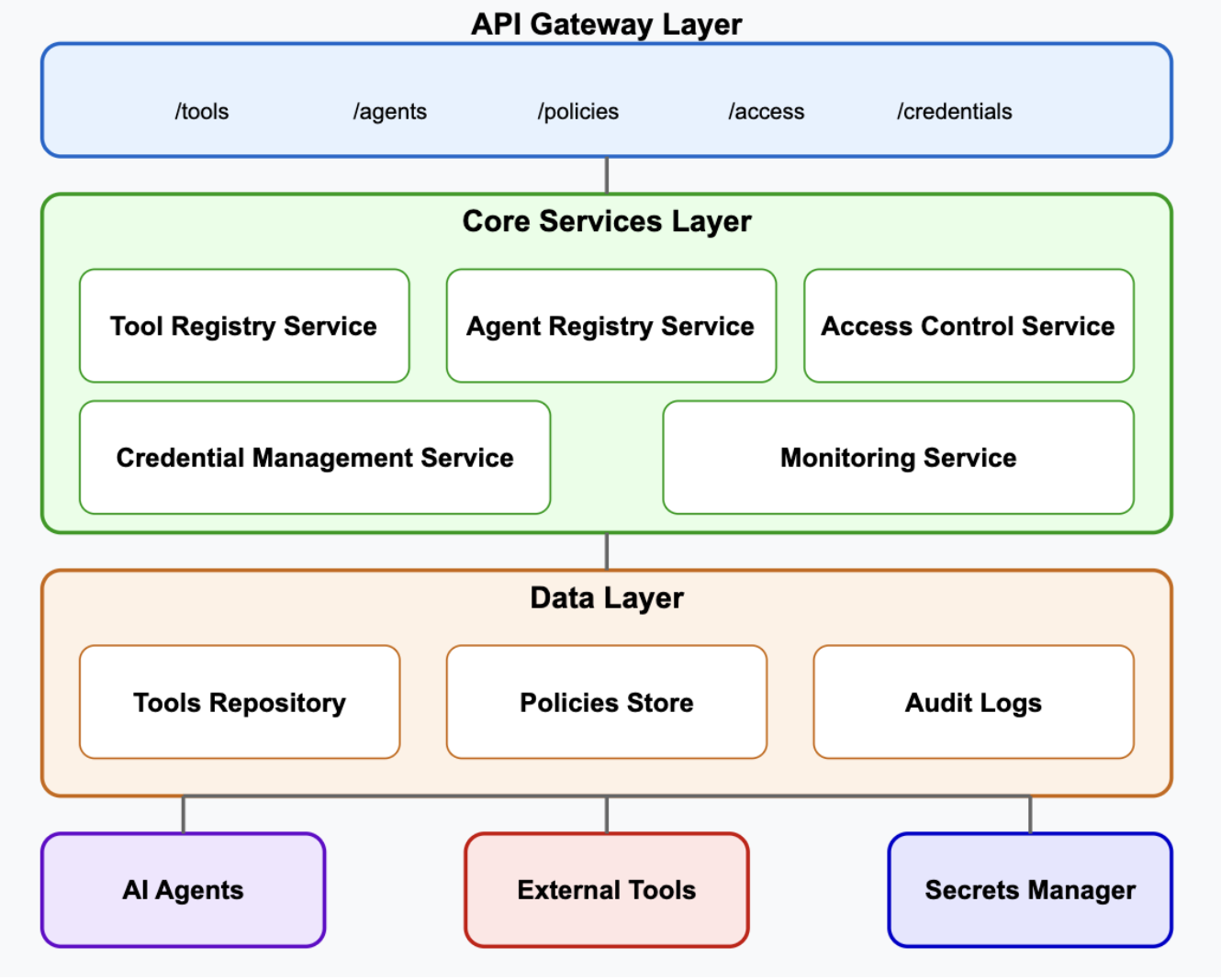} 

\caption{High level architecture of the Tool Registry} 
\label{fig:arch} 
\end{figure}

The system consists of five tightly integrated core components (See figure \ref{fig:arch}):
\begin{itemize}
    \item \textbf{Tool Registry Service:} Manages the registration, metadata storage, calculation, and maintenance of trust scores, and discovery queries for approved tools. The registration of the tool is performed by an administrator. 
    \item \textbf{Agent Registry Service:} Manages the registration, metadata storage, and identity verification for approved agents. Agent registration is also performed by an administrator. This service is crucial for linking authenticated requests to known, vetted agents.
    \item \textbf{Access Control Service:} Defines and enforces fine-grained policies that govern which registered agents can access which registered tools under specific conditions. Policies reference registered agent and tool identifiers.
    \item \textbf{Credential Management Service:} Securely provision, manage and revoke authentication credentials (ideally JIT) for authorized tool access requests originating from authenticated registered agents.
    \item \textbf{Monitoring Service:} Tracks and audits tools/agent registration events, discovery attempts, access requests, trust score changes, and usage patterns for security analysis and plugs into SOARs or SIEMs \cite{chandola2009anomaly}. 
\end{itemize}

\subsection{Architecture and Data Management} 
The framework employs a layered architecture, often implemented using microservices:
\begin{itemize}
    \item \textbf{API Gateway Layer:} Provides standardized RESTful endpoints for programmatic interaction (e.g., \texttt{/tools}, \texttt{/agents}, \texttt{/policies}, \texttt{/access}, \texttt{/monitor}). Access to registration endpoints (\texttt{/tools}, \texttt{/agents}) should be restricted to administrators.
    \item \textbf{Core Services Layer:} Implements the business logic for each of the five core components (Tool Registry, Agent Registry, Access Control, Credentials, Monitoring).
    \item \textbf{Data Layer:} Persists registry data, policies, logs, and configuration securely, often using a relational database like PostgreSQL.
\end{itemize}

\textbf{Key Data Stored:}
\begin{itemize}
    \item \textit{Tool Registry Service:} Stores comprehensive metadata about each administrator-registered tool. This includes a unique \texttt{tool\_id}, name, description, endpoint details, authentication configuration (specifying method like API key/OAuth and how JIT credentials should be applied, e.g., header name), parameter specifications, tool version, declared dependencies (e.g., required SDKs and their versions, potentially formatted as a Software Bill of Materials (SBOM)), current trust score (e.g., numeric scale 0-100 or categorical Low/Medium/High), last trust score calculation timestamp, owner, tags, and related policy identifiers. Crucially, sensitive credentials like static API keys for the tools themselves are typically not stored directly here but managed via integration with a secure secrets manager \cite{vault}, referenced in the \texttt{auth\_config}. 
    \item \textit{Agent Registry Service:} Stores detailed information about each administrator-registered agent. This includes a unique \texttt{agent\_id} (e.g., UUID), name, description, the administrator or entity responsible (creator), creation/update timestamps, assigned roles, a list of \texttt{allowed\_tools} (representing the potential tools the agent might be permitted to use, subject to policy), and potentially usage counters (\texttt{request\_count}).
    \item \textit{Access Control Service:} Stores policy definitions, linking principals (like agent IDs or roles from the Agent Registry) to resources (tool IDs from the Tool Registry) and permitted actions (\texttt{allowed\_scopes}). Policies include conditions (like rate limits, time of day), rules, priority, activation status, timestamps, and creator information.
\end{itemize}

\subsection{Agent and Tool Registration Process} 
The security of the framework hinges on a controlled registration process managed by trusted enterprise administrators:
\begin{enumerate}
    \item \textbf{Tool Registration:} An administrator registers a new tool via a secure interface or API call to the \texttt{/tools} endpoint. They provide the necessary metadata (name, description, endpoint, auth config, parameters, etc.). The Tool Registry Service validates the input and stores the verified tool information. This prevents arbitrary tool publication.
    \item \textbf{Agent Registration:} Similarly, an administrator registers a new agent via the \texttt{/agents} endpoint. They provide the agent's details (name, description, roles, initial list of \texttt{allowed\_tools}). The Agent Registry Service generates a unique \texttt{agent\_id} and stores the agent profile.
    \item \textbf{Initial Credential/Authentication:} Upon registration, the agent needs a mechanism to authenticate itself to the registry framework's API Gateway (e.g., receiving an initial API key or JWT tied to its \texttt{agent\_id}). This initial credential allows the agent to make authenticated calls to endpoints like \texttt{/access}.
    \item \textbf{Verification:} The primary verification step is the administrative action of registration itself. Only tools and agents explicitly approved and registered by the administrator are considered legitimate within the system. Further automated checks (like endpoint validation for tools) can supplement this.
    \item \textbf{Linking Identity to Requests:} When a registered agent makes a request (e.g., to \texttt{/access} for tool invocation), it must authenticate using its credential (e.g., presenting its JWT). The API Gateway or backend services validate this credential, extract the authenticated \texttt{agent\_id}, and use this ID to query the Agent Registry Service (to confirm the agent is valid and retrieve its profile, like roles, \texttt{allowed\_tools}) and the Access Control Service (to evaluate policies applicable to this specific \texttt{agent\_id}). This securely links the incoming request to a verified, registered agent identity.
\end{enumerate}

\begin{figure*}[!t]
\centering
\includegraphics[width=0.96\textwidth]{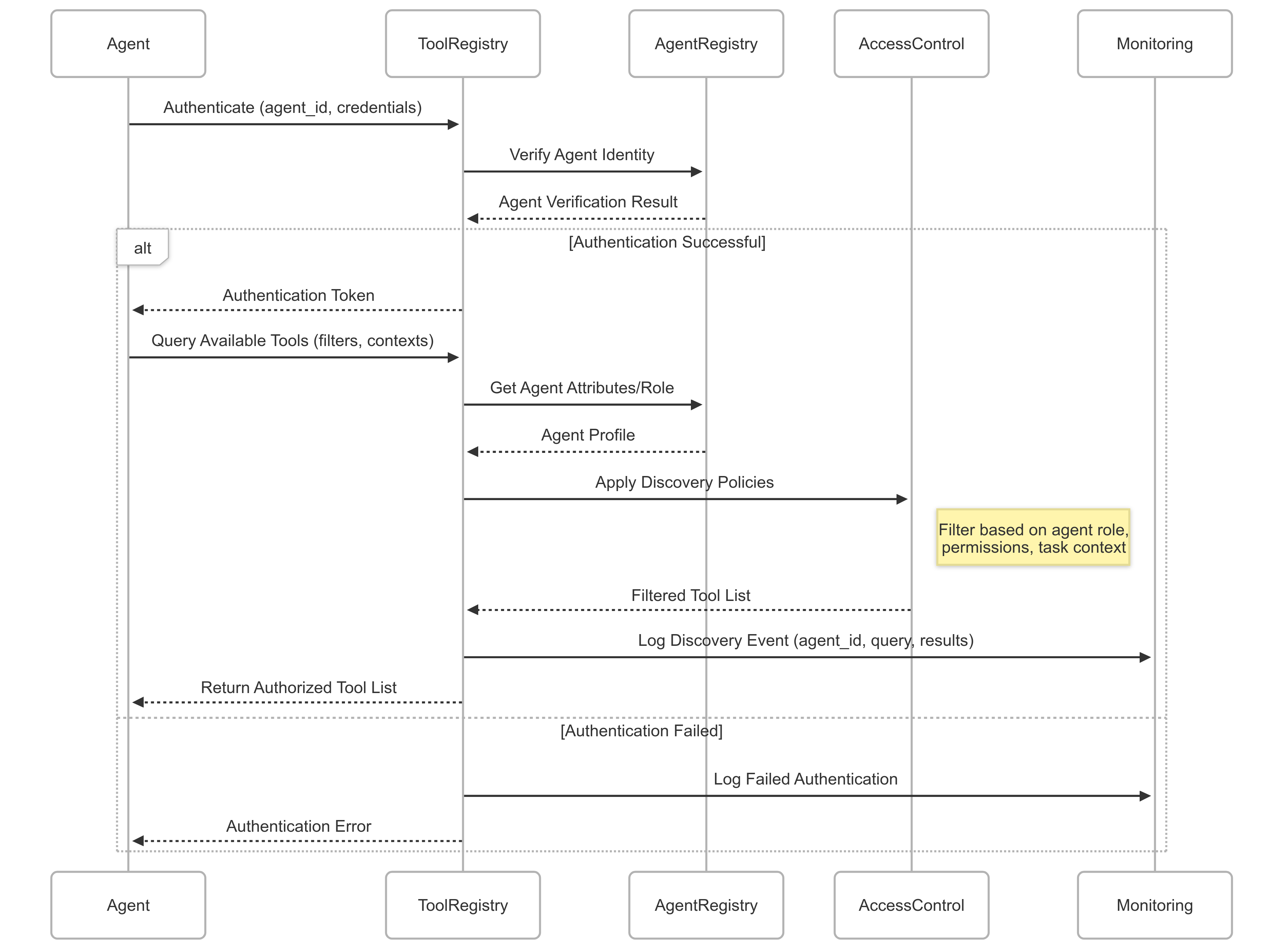} 
\caption{Discovery flow process to find tools} 
\label{fig:discovery_workflow} 
\end{figure*}

\section{Addressing Tool Squatting with the Registry Framework} 
\subsection{Preventing Deceptive Representation and Discovery Manipulation} 
The framework addresses the core of tool squatting (deceptive representation) and discovery manipulation through:
\begin{itemize}
    \item \textbf{Centralized, Admin-Verified Registration:} Only tools and agents vetted and explicitly registered by an enterprise administrator via the Tool and Agent Registry Services are listed. Malicious entities cannot register deceptive representations or inject themselves into the discovery process, effectively preventing spoofing.
    \item \textbf{Authenticated Discovery:} Agents must authenticate (using credentials tied to their registered \texttt{agent\_id}) with the registry framework before performing discovery queries via the \texttt{/tools} endpoint. This links discovery requests to verified identities.
    \item \textbf{Digitally Signed Metadata:} Agent Cards or equivalent discovery information can be cryptographically signed by the registry framework upon generation/query, allowing clients to verify the authenticity of the representation \cite{turner2011digital}.
    \item \textbf{Revocation Mechanisms:} Administrators can deactivate or delete registrations for compromised tools or agents, immediately invalidating their verified representation and associated credentials.
\end{itemize}
This ensures that agents discover authentic tools based on administrator-verified information, directly mitigating the risk of interacting with deceptively represented tools or agents.

\subsection{Securing Tool Discovery and Access} 
The registry framework secures the tool discovery and subsequent access process (See Figure \ref{fig:discovery_workflow}):
\begin{itemize}
    \item \textbf{Policy-Based Discovery Filtering:} Discovery results from the \texttt{/tools} endpoint are filtered based on the requesting agent's identity (obtained via authentication) and the policies defined in the Access Control Service. Agents only see the tools they are potentially authorized to access.
    \item \textbf{Capability Verification:} The Tool Registry stores verified tool metadata, allowing the framework (or the requesting agent) to validate that the capabilities advertised match registered specifications.
    \item \textbf{Enforced Access Control:} Before invoking a tool, the authenticated agent must request access permission via the \texttt{/access} endpoint. The Access Control Service evaluates policies based on the verified \texttt{agent\_id}, requested \texttt{tool\_id}, scope, and context, granting access only if allowed by policy and if the tool is listed in the agent's \texttt{allowed\_tools} profile.
\end{itemize}

\subsection{Just-in-Time Credential Provisioning} 
\begin{figure*}[!t]
\centering
\includegraphics[width=0.95\textwidth]{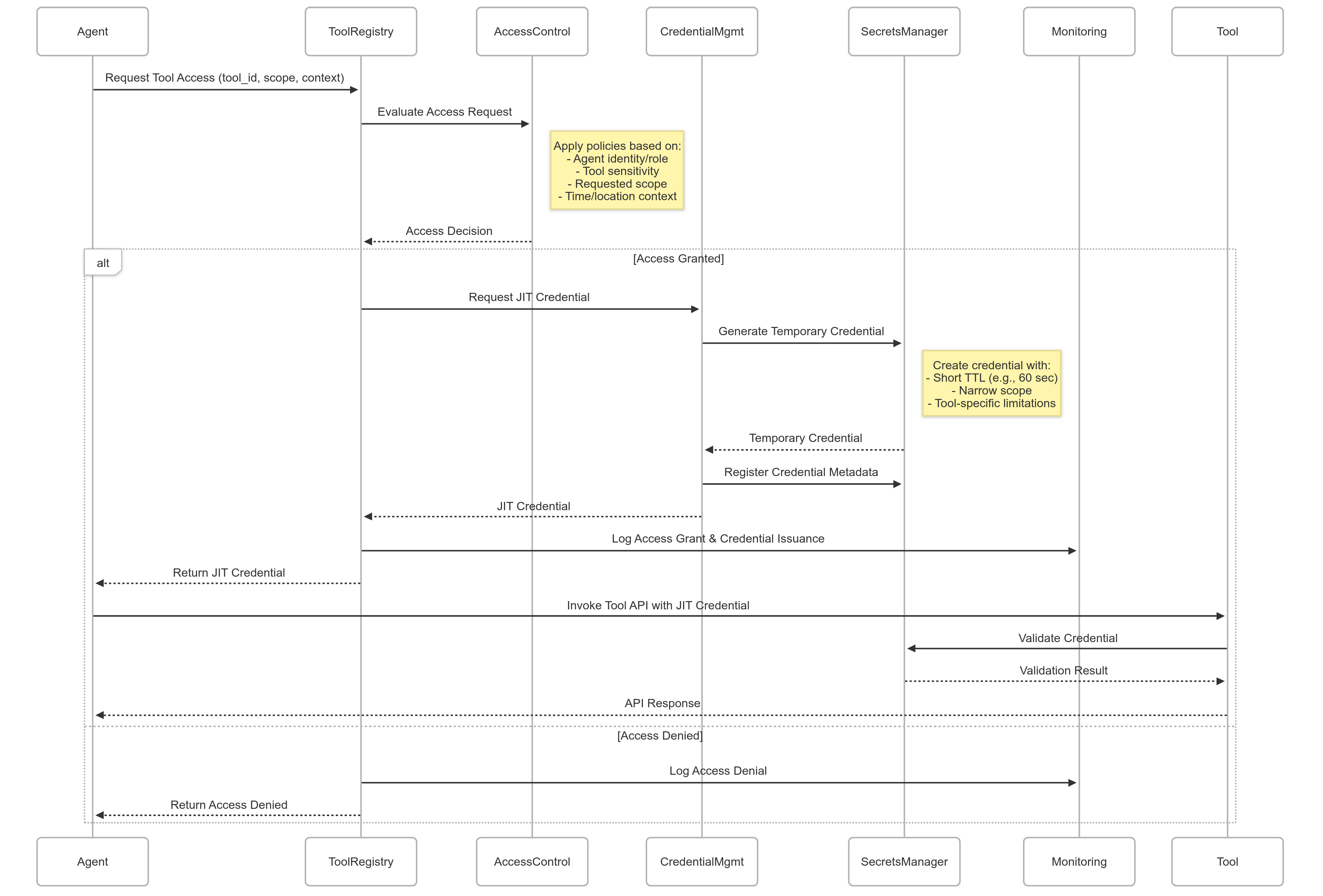} 

\caption{Invocation of authorized tools using JIT tokens} 
\label{fig:JIT_flow} 
\end{figure*}

A key innovation enabled by the registry framework is just-in-time (JIT) credential provisioning \cite{mahadeva2021jit} (See Figure \ref{fig:JIT_flow}):
\begin{itemize}
    \item \textbf{Ephemeral Credentials:} Instead of relying on long-lived, static credentials, the Credential Management Service issues temporary, short-lived tokens (e.g., JWTs, OAuth access tokens) upon successful authorization via the \texttt{/access} endpoint \cite{okta_jwt, ietf_oauth2}.
    \item \textbf{Scope Limitation:} These temporary credentials, issued to an authenticated and authorized agent, are narrowly scoped to the specific registered tool, operation, and potentially the data involved in the request.
    \item \textbf{Automatic Expiration/Revocation:} Credentials expire automatically. The registry framework can also actively revoke them if needed (e.g., if the agent or tool registration is deactivated).
    \item \textbf{Contextual Authorization:} Access decisions (policy evaluation) and credential issuance incorporate real-time context, such as the verified agent identity, task priority, system state, or detected threat levels.
\end{itemize}
This JIT mechanism significantly reduces the risk associated with credential theft or leakage and also follows the principle of least privilege\cite{mahadeva2021jit}.

\subsection{Monitoring and Anomaly Detection} 
The framework's monitoring service provides crucial visibility:
\begin{itemize}
    \item \textbf{Usage Logging:} All significant events are logged: tool/agent registration/modification by admins, discovery attempts, access requests (granted/denied based on policy evaluation for the specific \texttt{agent\_id} and \texttt{tool\_id}), and credential issuance events.
    \item \textbf{Pattern Analysis:} Logs can be fed into SIEM systems \cite{bace2001intrusion} or analyzed using AI/ML techniques \cite{chandola2009anomaly} to detect anomalous patterns indicative of misuse or attack (e.g., an agent suddenly requesting access to tools outside its normal profile, even if listed in \texttt{allowed\_tools}).
    \item \textbf{Rate Limiting:} The framework can enforce usage quotas and rate limits defined in policies, preventing resource exhaustion.
    \item \textbf{Alerting System:} Automated alerts can notify administrators of suspicious activities, policy violations, or potential security incidents in near real-time.
\end{itemize}

\subsection{Enhancing Security with Configurable Trust Scores} 
This framework not only prevents forged registrations but also implements a dynamic 'Trust Score' as a defense-in-depth mechanism for continuous risk evaluation of registered tools. This score adds an essential layer of security filtering, enabling agents to make more informed decisions about tool usage based on objective criteria.

This can be achieved by the Tool Registry Service, which computes and maintains a dynamic 'Trust Score' for every registered tool, offering a measurable assessment of its reliability. This score is affected by the tool's version currency, the existence and severity (CVSS) of known vulnerabilities in its dependencies (determined through SBOMs and CVE databases), the vendor's patching history, indications of ongoing exploits, and any upgrades. Scores are frequently adjusted upon updates to vulnerability data or tool dependencies, employing techniques that range from basic rule-based formulas to weighted models. Agents may filter tool discovery requests utilizing a minimal trust score parameter or employ the score to guide their selection among appropriate tools, balancing functionality with risk.

\section{Implementation Concept} 
To illustrate the framework's feasibility, we outline a potential proof-of-concept implementation as a microservices-based system utilizing common technologies:
\begin{itemize}
    \item \textbf{API Framework:} FastAPI (Python) for the RESTful API layer \cite{fastapi}.
    \item \textbf{Database:} PostgreSQL with SQLAlchemy ORM for persisting tool/agent metadata, policies, and logs.
    \item \textbf{Authentication:} JSON Web Tokens (JWT) for securing API endpoints and authenticating registered agents (carrying the \texttt{agent\_id}) \cite{okta_jwt}. Administrative access uses separate authentication.
    \item \textbf{Caching/Rate Limiting:} Redis for caching policies/metadata and implementing rate limiting \cite{redis}.
\end{itemize}
The system would expose key API endpoints:
\begin{itemize}
    \item \texttt{/tools}: (Admin restricted for CUD ops) Register, update, delete tools. (Agent accessible for R ops) Discover and retrieve details of authorized tools.
    \item \texttt{/agents}: (Admin restricted) Register, update, delete agents and manage their profiles (including \texttt{allowed\_tools}).
    \item \texttt{/policies}: (Admin restricted) Define, update, delete, and list access control policies referencing registered agent/tool IDs.
    \item \texttt{/credentials}: (Internal service endpoint) Manage the underlying credential issuance mechanisms, interacting with secrets management \cite{vault}.
    \item \texttt{/access}: (Agent accessible) Request access to a tool, triggering authentication, policy evaluation based on agent/tool IDs, and potential JIT credential issuance.
\end{itemize}

\subsection{Policy Definition Example} 
Policies stored by the Access Control Service use a flexible JSON schema, referencing registered agent principals and tool IDs. This example grants specific agents/roles access with rate limiting and a minimum trust score: 

\begin{lstlisting}[caption={Example Access Policy JSON}, label={lst:policy_json}]
{
  "policy_id": "pol_basic_summarizer_access",
  "name": "Basic Summarizer Access",
  "description": "Allows read/execute on Summarizer tool...",
  "tool_id": "tool_text_summarizer_v1", // References registered tool
  "principals": ["role:analyst", "agent:agent_id_123"], // References registered agents/roles
  "allowed_scopes": ["read", "execute"],
  "conditions": {
    "rate_limit": { "requests": 1000, "interval": "day" },
    "time_of_day": { "start": "09:00", "end": "17:00", "timezone": "UTC" },
    "minimum_tool_trust_score": 75 // Added minimum trust score condition
  },
  "rules": { "require_approval": false, "log_level": "INFO" },
  "priority": 10,
  "is_active": true
  // Timestamps and creator info also stored
}
\end{lstlisting}

\subsection{Tool Registration Example} 
Tools are registered by administrators with detailed metadata via the Tool Registry Service, including dependency and trust score information: 

\begin{lstlisting}[caption={Example Tool Registration JSON}, label={lst:tool_reg_json}]
{
  "tool_id": "tool_text_summarizer_v1", // Assigned by registry or admin
  "name": "Text Summarizer",
  "description": "Summarizes long text documents...",
  "api_endpoint": "https://api.example.com/summarize/v1",
  "auth_method": "api_key", // Or "oauth", "jwt" etc.
  "auth_config": { // How the tool expects auth
    "mechanism": "header",
    "header_name": "X-API-Key",
    "credential_placeholder": "${JIT_API_KEY}" // Placeholder for JIT injection
    // May include references to secrets in a vault [25]
  },
  "parameters": { /* ... input/output schema ... */ },
  "version": "1.0.0",
  "dependencies": [ // Added dependencies
    {"component": "summarization-sdk", "version": "2.5.0"},
    {"component": "logging-lib", "version": "4.0.0"}
  ],
  "trust_score": 92, // Added trust score
  "last_trust_score_update": "2025-04-21T18:00:00Z", // Added timestamp
  "owner": "nlp-team@example.com",
  "tags": ["text", "summarization", "nlp"]
  // Associated policy IDs might also be linked here
}
\end{lstlisting}

\subsection{Just-in-Time Credential Flow} 
The JIT credential flow operates conceptually as follows \cite{mahadeva2021jit}:
\begin{enumerate}
    \item An agent, authenticated via JWT containing its registered \texttt{agent\_id}, requests access to \texttt{tool\_text\_summarizer\_v1} with scope \texttt{execute} via the \texttt{/access} endpoint.
    \item The framework verifies the agent's JWT and extracts the \texttt{agent\_id}.
    \item The Agent Registry Service is consulted to confirm \texttt{agent\_id} is valid and retrieve its profile (e.g., roles, \texttt{allowed\_tools} list). Assume \texttt{tool\_text\_summarizer\_v1} is in \texttt{allowed\_tools}.
    \item The Access Control Service retrieves relevant policies based on the \texttt{agent\_id}, its roles, the requested \texttt{tool\_id}, and scope.
    \item Policies (like the example in 6.1) are evaluated against the request context. This now includes checking the tool's current trust score against the policy's minimum requirement. 
    \item If access is granted (policy allows, conditions met, trust score sufficient), the request is forwarded to the Credential Management Service.
    \item The Credential Management Service generates a short-lived credential (e.g., an API key fetched from a vault or a scoped JWT) based on the tool's \texttt{auth\_config}.
    \item The temporary credential is returned to the agent.
    \item The agent uses the credential to call the tool's API endpoint.
    \item The credential expires automatically or can be revoked.
    \item All steps are logged by the Monitoring Service with \texttt{agent\_id}, \texttt{tool\_id}, policy details, etc.
\end{enumerate}

\begin{figure*}[!t]
\centering
\includegraphics[width=0.95\textwidth]{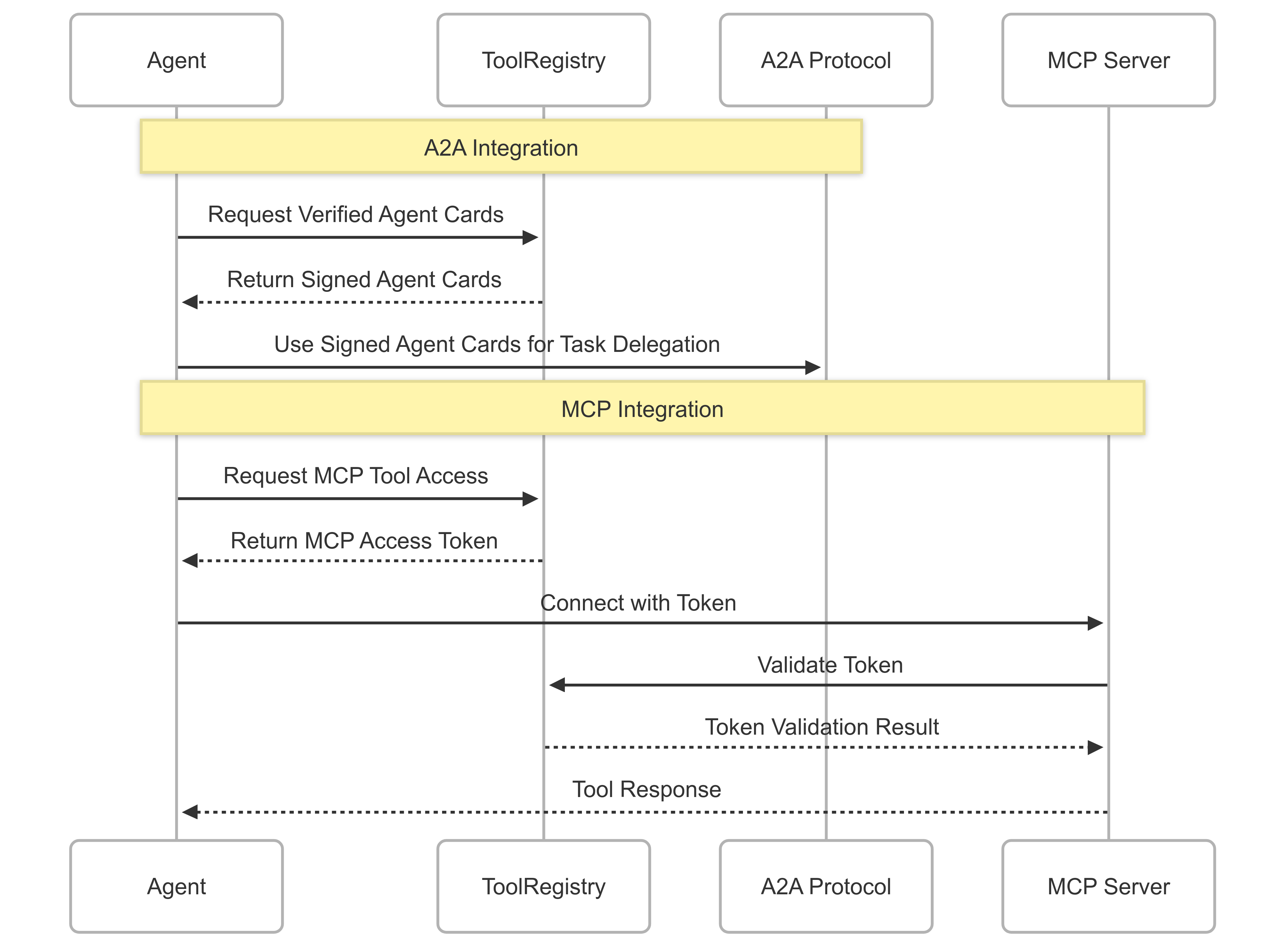} 

\caption{High level flow showing MCP and A2A Integration} 
\label{fig:mcp_a2a_flow} 
\end{figure*}

\section{Discussion and Future Work} 
\subsection{Security vs. Usability Trade-offs} 
Our framework design highlights inherent trade-offs:
\begin{itemize}
    \item \textbf{Administrative Control vs. Agility:} Requiring administrator registration for all agents and tools significantly enhances security and prevents squatting but introduces a potential bottleneck and reduces the agility compared to open registration systems.
    \item \textbf{Policy Granularity vs. Complexity:} Fine-grained policies provide robust security but increase management complexity.
    \item \textbf{Performance Impact:} Security checks introduce latency. Optimizations are needed \cite{redis}, although specific performance evaluation is outside the scope of this framework description.
\end{itemize}

\subsection{Integration with A2A and MCP Protocols} 
The Tool Registry framework complements A2A \cite{google2025a2a, google2025a2a_github} and MCP \cite{anthropic2024mcp_report, narajala2025enterprise} (See Figure \ref{fig:mcp_a2a_flow}).
\\
\textbf{A2A Integration:}
\begin{itemize}
    \item \textit{Trusted Agent Card Source:} The framework acts as the source for Agent Cards, generated only for admin-registered agents/tools.
    \item \textit{Secured Discovery:} A2A discovery queries are routed via the framework's API, applying authentication and policy filtering.
    \item \textit{Authorized Task Delegation:} A2A task delegation requires an \texttt{/access} check against the framework to confirm authorization for the specific agent-tool interaction and potentially receive a JIT token.
\end{itemize}
\textbf{MCP Integration:}
\begin{itemize}
    \item \textit{Secure MCP Server Registration:} MCP servers (representing tools) must be registered by an admin (or multi-part approval process) in the Tool Registry.
    \item \textit{Authenticated Connection:} MCP clients (agents) authenticate to MCP servers using JIT credentials obtained via the framework's \texttt{/access} endpoint after verifying the agent's registration and policies.
    \item \textit{Server-Side Authorization:} MCP servers can validate presented JIT tokens or query the framework's Access Control Service to authorize requests based on the verified client \texttt{agent\_id}.
    \item \textit{Client-Side Verification:} MCP clients can verify MCP server authenticity against the Tool Registry.
\end{itemize}

\subsection{Limitations of the Tool Registry Framework} 
\begin{itemize}
    \item \textbf{Centralization Bottleneck/Single Point of Failure:} The core services are centralized. High availability and scaling are crucial.
    \item \textbf{Administrative Overhead:} Reliance on administrators for all registrations can be a bottleneck, especially in large, dynamic environments. We are assuming we always have a trusted administrator.
    \item \textbf{Complexity:} Implementing and managing the framework, policies, and integrations is complex. Tool compatibility with JIT/scoped credentials may vary.
    \item \textbf{Scope of Protection:} Primarily secures discovery and access initiation. Doesn't protect against tool vulnerabilities or fully trusted compromised agents (though JIT/monitoring/trust scores help). 
    \item \textbf{Verification Rigor:} Security depends on the administrator's diligence during registration and the robustness of the agent authentication mechanism.
\end{itemize}

\subsection{Enterprise Deployment Considerations} 
\begin{itemize}
    \item \textbf{Centralized Governance Value:} Essential for enterprise control, compliance, audit trails \cite{bace2001intrusion}, and fine-grained access management beyond basic protocol capabilities.
    \item \textbf{Integration with Existing Infrastructure:} Crucial to integrate with IAM \cite{idsa_iam}, secrets management, SIEM, etc. Agent registration might leverage existing IAM identities.
    \item \textbf{Verification Process Definition:} While admin (or multi-part approval process) registration is key, define supplementary checks (code scans, ownership validation) as part of the admin's workflow. Define clear processes for agent/tool onboarding and offboarding.
    \item \textbf{Policy Lifecycle Management:} Use "policy as code" \cite{opa_policy}, establish review workflows, and potentially leverage roles defined in the Agent Registry for simpler policy definition.
    \item \textbf{Operational Resilience:} Design for high availability, scalability, and disaster recovery.
    \item \textbf{Monitoring and Alerting:} Implement comprehensive monitoring and alerts integrated with security operations \cite{bace2001intrusion}.
\end{itemize}

\subsection{Future Directions} 
\begin{itemize}
    \item \textbf{Automated Verification:} Exploring ways to automate parts of the agent/tool verification process managed by administrators.
    \item \textbf{Decentralized Trust:} Investigating DIDs/VCs \cite{w3c_vc} to potentially reduce reliance on central admin control for certain trust aspects.
    \item \textbf{Standardization Efforts:} Contributing these concepts to evolving standards.
    \item \textbf{Cross-Platform Interoperability:} Designing mechanisms for secure interaction across federated registries.
    \item \textbf{Empirical Evaluation:} Simulate tool-squatting attacks (both external and insider-actor scenarios) to quantify prevention/detection rates and perform red teaming on the Registry. 
    \item \textbf{Formalization of Trust-Score Mechanism:} Define the algorithm, weights, and other details for the trust score mechanism. 
\end{itemize}

\section{Conclusion} 
The proliferation of GenAI multi-agent systems introduces powerful capabilities but also significant security challenges like tool squatting (the deceptive representation of tools). Our proposed Tool Registry framework, centered on administrator-controlled registration via dedicated Agent and Tool Registry services, offers a structured and implementable solution framework. By incorporating centralized verification through administrative oversight, fine-grained policy-based access control linked to registered identities, dynamic trust scoring, just-in-time credential provisioning, and comprehensive monitoring, the framework establishes a secure foundation for managing agent-tool interactions within an enterprise. While acknowledging limitations like administrative dependency and implementation complexity, this framework offers a practical approach to securing the critical interface between agents and external tools, enabling organizations to harness MAS capabilities while maintaining a strong security posture. Further work would involve detailed implementation and empirical evaluation to validate the framework's effectiveness in practice. 

\bibliographystyle{IEEEtran}
\bibliography{IEEEabrv,references} 

\begin{thebibliography}{10}
\providecommand{\url}[1]{#1}
\csname url@samestyle\endcsname
\providecommand{\newblock}{\relax}
\providecommand{\bibinfo}[2]{#2}
\providecommand{\BIBentrySTDinterwordspacing}{\spaceskip=0pt\relax}
\providecommand{\BIBentryALTinterwordstretchfactor}{4}
\providecommand{\BIBentryALTinterwordspacing}{\spaceskip=\fontdimen2\font plus
\BIBentryALTinterwordstretchfactor\fontdimen3\font minus \fontdimen4\font\relax}
\providecommand{\BIBforeignlanguage}[2]{{%
\expandafter\ifx\csname l@#1\endcsname\relax
\typeout{** WARNING: IEEEtran.bst: No hyphenation pattern has been}%
\typeout{** loaded for the language `#1'. Using the pattern for}%
\typeout{** the default language instead.}%
\else
\language=\csname l@#1\endcsname
\fi
#2}}
\providecommand{\BIBdecl}{\relax}
\BIBdecl

\bibitem{google2025a2a}
\BIBentryALTinterwordspacing
{Google Research}, ``{Agent2Agent: An Open Interoperability Protocol for AI Agents},'' Google, Tech. Rep., 2025. [Online]. Available: \url{https://developers.googleblog.com/en/a2a-a-new-era-of-agent-interoperability/}
\BIBentrySTDinterwordspacing

\bibitem{anthropic2024mcp_report}
\BIBentryALTinterwordspacing
Anthropic, ``{Model Context Protocol: A Standard Interface for AI Context Access},'' Anthropic, Tech. Rep., 2024. [Online]. Available: \url{https://docs.anthropic.com/en/docs/agents-and-tools/mcp}
\BIBentrySTDinterwordspacing

\bibitem{mas_threat_model_2025}
\BIBentryALTinterwordspacing
K.~Huang, A.~Sheriff, J.~Sotiropoulos, R.~F. Del, and V.~Lu, ``Multi-agentic system threat modelling guide {OWASP} {GenAI} security project,'' Apr. 2025. [Online]. Available: \url{https://www.researchgate.net/publication/391204915_Multi-Agentic_system_Threat_Modelling_Guide_OWASP_GenAI_Security_Project}
\BIBentrySTDinterwordspacing

\bibitem{llm_genai_security_2025}
\BIBentryALTinterwordspacing
E.~G. Junior, S.~Clinton, C.~Hughes, V.~S. Narajala, and T.~Holmes, ``{LLM} and {GenAI} data security best practices,'' Feb. 2025. [Online]. Available: \url{https://www.researchgate.net/publication/391204648_LLM_and_GenAI_Data_Security_Best_Practices}
\BIBentrySTDinterwordspacing

\bibitem{han2024llmagent}
\BIBentryALTinterwordspacing
S.~Han, Q.~Zhang, Y.~Yao, W.~Jin, Z.~Xu, and C.~He, ``{LLM Multi-Agent Systems: Challenges and Open Problems},'' \emph{arXiv preprint arXiv:2402.03578}, 2024. [Online]. Available: \url{https://arxiv.org/pdf/2402.03578}
\BIBentrySTDinterwordspacing

\bibitem{mahadeva2021jit}
N.~Mahadeva and K.~Tuma, ``{Just-In-Time Access: Reducing the Blast Radius of Compromised Credentials},'' \emph{USENIX ;login:}, vol.~46, no.~4, pp. 6--11, 2021.

\bibitem{google2025a2a_github}
\BIBentryALTinterwordspacing
GitHub, ``{Google/A2A: An open protocol enabling communication between opaque agentic applications},'' GitHub Repository, 2025. [Online]. Available: \url{https://github.com/google/A2A}
\BIBentrySTDinterwordspacing

\bibitem{narajala2025enterprise}
\BIBentryALTinterwordspacing
V.~S. Narajala and I.~Habler, ``{Enterprise-Grade Security for the Model Context Protocol (MCP): Frameworks and Mitigation Strategies},'' \emph{arXiv preprint arXiv:2504.08623}, 2025. [Online]. Available: \url{https://arxiv.org/abs/2504.08623}
\BIBentrySTDinterwordspacing

\bibitem{he2025redteaming}
\BIBentryALTinterwordspacing
P.~He, Y.~Lin, S.~Dong, H.~Xu, Y.~Xing, and H.~Liu, ``{Red-Teaming LLM Multi-Agent Systems via Communication Attacks},'' \emph{arXiv preprint arXiv:2502.14847}, 2025. [Online]. Available: \url{https://arxiv.org/abs/2502.14847}
\BIBentrySTDinterwordspacing

\bibitem{triedman2025multiagent}
\BIBentryALTinterwordspacing
H.~Triedman, R.~Jha, and V.~Shmatikov, ``{Multi-Agent Systems Execute Arbitrary Malicious Code},'' \emph{arXiv preprint arXiv:2503.12188}, 2025. [Online]. Available: \url{https://arxiv.org/pdf/2503.12188v1}
\BIBentrySTDinterwordspacing

\bibitem{mccarthy2025mcp}
\BIBentryALTinterwordspacing
R.~McCarthy, ``{Research Briefing: MCP Security},'' Wiz Blog, April 2025. [Online]. Available: \url{https://www.wiz.io/blog/mcp-security-research-briefing}
\BIBentrySTDinterwordspacing

\bibitem{zeng2019comprehensive}
Y.~Zeng, T.~Zang, Y.~Zhang, X.~Chen, and Y.~Wang, ``{A Comprehensive Measurement Study of Domain-Squatting Abuse},'' in \emph{IEEE International Conference on Communications (ICC)}, 2019, pp. 1--6.

\bibitem{securing_a2a}
\BIBentryALTinterwordspacing
I.~Habler, K.~Huang, V.~S. Narajala, and P.~Kulkarni, ``Building a secure agentic {AI} application leveraging {A2A} protocol,'' 2025. [Online]. Available: \url{https://www.arxiv.org/abs/2504.16902}
\BIBentrySTDinterwordspacing

\bibitem{chandola2009anomaly}
V.~Chandola, A.~Banerjee, and V.~Kumar, ``{Anomaly detection: A survey},'' \emph{ACM Computing Surveys (CSUR)}, vol.~41, no.~3, pp. 1--58, 2009.

\bibitem{vault}
\BIBentryALTinterwordspacing
{HashiCorp}, ``{HashiCorp Vault},'' Web Page. [Online]. Available: \url{https://www.vaultproject.io/}
\BIBentrySTDinterwordspacing

\bibitem{turner2011digital}
D.~Turner and T.~Polk, Eds., \emph{{Digital Signatures}}, 6th~ed.\hskip 1em plus 0.5em minus 0.4em\relax Wiley, 2011, vol.~1.

\bibitem{okta_jwt}
\BIBentryALTinterwordspacing
{Okta Developer}, ``{An Introduction to JSON Web Tokens},'' Okta Developer Documentation. [Online]. Available: \url{https://developer.okta.com/docs/concepts/json-web-tokens/}
\BIBentrySTDinterwordspacing

\bibitem{ietf_oauth2}
{IETF}, ``{RFC 6749: The OAuth 2.0 Authorization Framework},'' Internet Engineering Task Force, 2012.

\bibitem{bace2001intrusion}
R.~Bace and P.~Mell, ``{Intrusion Detection Systems},'' National Institute of Standards and Technology, Tech. Rep. Special Publication 800-31, 2001.

\bibitem{fastapi}
\BIBentryALTinterwordspacing
FastAPI, ``{FastAPI Web Framework},'' Web Page. [Online]. Available: \url{https://fastapi.tiangolo.com/}
\BIBentrySTDinterwordspacing

\bibitem{redis}
\BIBentryALTinterwordspacing
Redis, ``{Redis In-Memory Data Structure Store},'' Web Page. [Online]. Available: \url{https://redis.io/}
\BIBentrySTDinterwordspacing

\bibitem{idsa_iam}
\BIBentryALTinterwordspacing
{Identity Defined Security Alliance (IDSA)}, ``{What is Identity and Access Management (IAM)?}'' Web Page. [Online]. Available: \url{https://www.idsalliance.org/what-is-identity-and-access-management-iam/}
\BIBentrySTDinterwordspacing

\bibitem{opa_policy}
\BIBentryALTinterwordspacing
{Open Policy Agent (OPA)}, ``{Policy as Code},'' Web Page. [Online]. Available: \url{https://www.openpolicyagent.org/docs/latest/policy-as-code/}
\BIBentrySTDinterwordspacing

\bibitem{w3c_vc}
\BIBentryALTinterwordspacing
{W3C Credentials Community Group}, ``{Verifiable Credentials Data Model v1.1},'' W3C Recommendation, 2022. [Online]. Available: \url{https://www.w3.org/TR/vc-data-model/}
\BIBentrySTDinterwordspacing

\end{thebibliography}

\end{document}